preprint {WIS-01/22 Nov-DPP}
\documentstyle[prc,aps,epsf,amssymb]{revtex}

\newcommand{\bmi}{{\mbox{\boldmath $i$}}}

\newcommand{\bmr}{{\mbox{\boldmath $r$}}}

\newcommand{\bmq}{{\mbox{\boldmath $q$}}}
\newcommand{\bmb}{{\mbox{\boldmath $b$}}}

\newcommand{\qq}{{|\mbox{\boldmath $q$}|}}

\begin{document}
\preprint {WIS-01/22 Nov-DPP}
\draft
\date{\today}
\title{Description of inclusive scattering of 4.045 GeV electrons from $D$.}
\author{A.S. Rinat and M. F. Taragin}
\address{Weizmann Institute of Science, Department of Particle Physics,
Rehovot 76100, Israel}
\maketitle
\begin{abstract}

We exploit a relationship between the Structure Functions of nucleons, the
physical deuteron and of a deuteron, composed of point-nucleons to compute
angular distributions of inclusive cross sections of 4.05 GeV electrons. We
report general agreement with data and interpret the remaining
discrepancies. We discuss the potential of the data for information on neutron
structure functions $F_k^n(x,Q^2)$ and the static form factor $G_M^n(Q^2)$.
25.20.Fj,13.60Hb
\end{abstract}

\section{Introduction.}

In the early days of inclusive scattering experiments the deuteron
($D$) as a target was only second in importance to the proton \cite{lung,x}.
The obvious reason was and is the quest of information on the neutron, in
particular of its static form factors $G_{E,M}^n$ and the structure functions
$F_k^n$. Later experiments for ranges of fixed $Q^2$, and similar ones for
$^3$He, $^4$He, related mostly to the issue of scaling \cite{rock}.

Modern inclusive scattering
experiments \cite{day,arr89} have used targets with $A\ge 4$, and only
recently  have  data on double-differential cross sections on the $D$ for
a beam energy $E=4.045$ GeV become available \cite{arrd,nicu}. Those are
the first of its kind, and to our knowledge no calculation has thus far
been made. The incentive for such a calculation is two-fold. First, one can
compute the nuclear input with great precision, making for a stringent
comparison with data. Second, we shall emphasize the  unusual spatial
extension (or the nucleon momentum distribution) of the deuteron,
which expresses itself  in non-standard features in cross sections.
In turn those features can be exploited to  extract $G_M^n$ \cite{gmn}.

\section {Computation the inclusive cross sections on the $D$.}

We start with the expression for inclusive cross sections per nucleon of
unpolarized high-energy electrons from randomly oriented arbitrary nuclear
targets as function of the scattering angle $\theta$
and the energy loss of the beam $\nu$
\begin{eqnarray}
\frac{d^2\sigma^{eD}(E;\theta,\nu)/2}{d\Omega\,d\nu}
=\frac{2}{M}
\sigma_M(E;\theta,\nu)\bigg\lbrack\frac {xM^2}{Q^2}F_2^D(x,Q^2)+
{\rm tan}^2(\theta/2)F_1^D(x,Q^2)\bigg\rbrack\ ,
\label{a1}
\end{eqnarray}
$F_{1,2}^D(x,Q^2)$ are the Structure Functions (SF)
per nucleon of the $D$, which may be expressed
in terms of the squared 4-momentum transfer $Q^2=\bmq^2-\nu^2$ and the Bjorken
variable  $x$ with range $0\le x=Q^2/2M\nu\le 2$ ($M$ is the nucleon mass) .
For given beam energy $E$
the pairs $(\theta,\nu)$ and $(x,Q^2)$ are alternative kinematic variables.

We base calculations of inclusive cross sections on the following relation
between the  SF of the target $F^D$  and of nucleons $F^N$ \cite{gr}
\begin{eqnarray}
F_k^D(x,Q^2)=\int_x^2 dz
f^{PN,D}(z,Q^2)\bigg [F_k^p\bigg (\frac {x}{z},Q^2
\bigg )+ F_k^n\bigg (\frac {x}{z},Q^2\bigg )\bigg ]\bigg /2 \ ,
\label{a2}
\end{eqnarray}
with $F_k^{p,n}$ the nucleon SF and $f^{PN,A}$, the  SF of a nucleus composed
of point-nucleons. In the expression for $F_k^D$ for given $k$ one ought to
include coefficients, which mix different nucleon SF \cite{atw}. Their effect
decrease with increasing $Q^2$ and we suggest that those may be neglected
for the $D$ data under investigation.

Eqs. (\ref{a2}) describes parton degrees of freedom of nucleons
but not those, originating  from other  sources, for instance from
virtual bosons. The latter contributions, as well as anti-screening
effects decrease  with increasing $x$, limiting the use of Eq. (\ref{a2})
to $x\gtrsim$ 0.15-0.20 \cite{pion}, well below the
smallest $x$ reached in the data. Finally,  Eq. (\ref{a2}) has been
estimated to hold for $Q^2\gtrsim Q_c^2\approx$2-2.5 GeV$^2$
\cite{rt1,commar}.

For use below, we mention a separation of nuclear SF, and consequently
of cross sections into nucleon-elastic (NE) and nucleon-inelastic (NI)
components. Those correspond to contributions which, after absorption of
virtual photons, nucleons are  not (NE), or are excited (NI) \cite{commar}.

Eq. (\ref{a2}) is routinely used in the Plane Wave Impulse Approximation
(PWIA) (see for instance Ref. \onlinecite{ciof1}). Here
we adhere  to a non-perturbative version with on-shell nucleons
\cite{gr} and which in the past has been applied to nuclei with $A\ge 12$
\cite{rt1,rt11}.

There are several incentives to measure and to compute inclusive scattering
on light nuclei. For the time being, we recall that the nuclear
specificity of SF resides in $f^{PN,A}$, which is the SF of a nucleus,
composed of point-nucleons. In contrast to nuclei with  $A\ge 12$, for which
the nuclear part of $f^{PN,A}$ can only be computed approximately, for light
nuclei such a calculation can be performed with great  precision \cite{bench}.

A first description of data on a light nucleus $^4$He, exploiting a
relativistic version \cite{gr1} of the Gersch-Rodriguez-Smith (GRS) theory
for SF \cite{grs} has  been completed \cite{viv}. Below we shall  present
the $D$ case. For it, there exists only the single-$N$ density matrix
${\cal A}(\bmr,\bmr')$, which in cylindrical coordinates ($\bmr=\bmb,z$) is
diagonal in $\bmb$. With  $s=z-z'$ it reads
\begin{eqnarray}
{\cal A}(b,z;s)&=&\frac{1}{3}\sum_{M_D}\sum_{\sigma_1,\sigma_2}
\langle \Phi_{M_D}(\bmr-s\hat\bmi_q; \sigma_1,\sigma_2)
\Phi_{M_D}(\bmr; \sigma_1,\sigma_2)\rangle
\nonumber\\
&=&\frac{1}{4\pi rr'}[u(r)u(r')+w(r)w(r')]P_2(t)
\label{a5}
\end{eqnarray}
Above one sums over the direction of the spin of the unpolarized $D$
and integrates over nuclear spins (see for instance Ref. \cite{niko}).

The functions $u,w$ in Eq. (\ref{a5}) are the standard radial $L$=0,2
components of the $D$ ground state \cite{mach}. In Eq. ({\ref{a5})
$r=\sqrt{b^2+z^2)},\,r'=\sqrt{b^2+(z-s)^2}$ while  $P_2(t)=(3t^2-1)/2$,
with $t=(s^2-r^2-r'^2)/2rr'$. The latter corresponds to the choice of the
$z$-axis  along the direction of the 3-momentum transfer $\bmq$.

First we choose as kinematic variables the 3-momentum transfer $|\bmq| $and
a scaling variable $y$, which replaces the energy loss $\nu$.
In terms of those, we decompose the relativistic reduced structure function
as $\phi(q,y)=\phi_0(y)+\phi^{FSI}(q,y)$, which are the asymptotic limit and
the $q$-dependent Final State Interactions (FSI) which  perturbs the former
\cite{gr1}. Both employ  the above one-nucleon density ${\cal A}$
\cite{rt1}
\begin{mathletters}
\label{a6}
\begin{eqnarray}
\phi_0(q)&=&2\pi\int_{-\infty}^{\infty} \frac {ds}{2\pi} e^{iys}
\int_0^{\infty} db b \int_{\infty}^{\infty} dz {\cal A}(bz;s)
\label{a6a}\\
\frac {M}{\qq}\phi^{FSI}(q,y)&=&2\pi \int_{-\infty}^{\infty} \frac {ds}{2\pi}
\int_0^{\infty} db b \int_{-\infty}^{\infty} dz {\cal A}(bz;s)
[\tilde \Gamma_q(bz;s)-1]
\label{a6b}
\end{eqnarray}
\end{mathletters}
The FSI term for the $D$ contains the off-shell $pn$ scattering amplitude.
Eq. (\ref{a6b}) above uses its eikonal approximation, which in
coordinate representation  is proportional to the off-shell profile
$\tilde\Gamma_q(bz,s)$, in turn  approximately  related to its on-shell
analog \cite{rt1,rt2}. The latter may  in a standard way be
expressed in terms of elastic scattering observables
\begin{mathletters}
\label{a7}
\begin{eqnarray}
\tilde\Gamma_q(bz,s)&\approx&\bigg [1-s\frac {\partial}{\partial s}\bigg ]
\theta(z)\theta(s-z) \Gamma_q^{(1)}(b)
\label{a7a}\\
\Gamma_q^{(1)}(b)&\approx &\frac {1}{2}\sigma_q^{tot}[1-i\tau_q]\frac
{Q_0^2(q)}{4\pi} e^{-b^2Q_0^2/4}
\label{a7b}
\end{eqnarray}
\end{mathletters}
Substitution into Eq. (\ref{a6b}) gives for the FSI part
\begin{eqnarray}
\frac{M}{\qq}\phi^{FSI}(q,y)\approx 2\pi \int_{-\infty}^{\infty}
\frac {ds}{2\pi} \int_0^{\infty} db b \Gamma_q(b)
\bigg [\int_0^s dz {\cal A}_1(bz;s)-s{\cal A}_1(bs;s)\bigg ]
\label{a8}
\end{eqnarray}
It is through Eq. (\ref{a7a}) that $\phi^{FSI}(q,y)$ acquires model-dependence.

In previous applications for $A\ge 4$ \cite{rt1,rt11}, $y$ has been
taken to be a relativistic version of the West-GRS scaling variable for
$A\to \infty$ \cite{sag}
\begin{eqnarray}
y_G=\frac {M\nu}{\qq}\bigg \lbrack 1-\frac{\Delta}{M}-x \bigg \rbrack \ ,
\label{a9}
\end{eqnarray}
with $\Delta$ some average separation energy. The above expression disregards
for $A\ge 4$ the energy of the recoiling spectator, but for the $D$ this is
not accurate enough. Its inclusion leads to the replacement
\cite{gr1}
\begin{eqnarray}
y_G\to  y_G^D=M\frac {\qq}{\nu}\bigg [\sqrt{1+2\frac {\nu}{\qq}\frac
{y_G}{M}}-1\bigg ]
\label{a10}
\end{eqnarray}
In the end, one converts the structure function $\phi^D(q,y)$ into a
dimensionless equivalent in terms of $x,Q^2$ \cite{rt1,rt11} as required in
Eq. (\ref{a2})
\begin{eqnarray}
f^{PN,D}(x,Q^2)=\bigg |\frac{\partial  y_G^D}{\partial x}\bigg |
\phi^D\bigg (q(x,Q^2),y_G^D(x,Q^2) \bigg )
\label{a11}
\end{eqnarray}
In order to obtain $F_k^D$,  the above has to be folded into $F_k^{p,n}$
(cf. Eq. (\ref{a2})). Regarding the latter, there are data
on $F_2^p$ \cite{arn} and less accurate older ones for $F_1^p$
\cite{bod}. Both do not reach the elastic region $x\lesssim 1$ but
parametrizations cover the entire $x$-range. With no direct information on the
neutron SF $F_k^n$, one usually assumes the $'$primitive$'$ choice $F_k^n=
2F_k^D-F_k^p$ \cite{bod,arn1} which corresponds to free $p,n$ in the $D$.

In Fig. 1 we display total $D$ cross sections per nucleon (\ref{a1})
and their NE parts for inclusive scattering of  $E$=4.045 GeV  electrons
as function of the scattering angles
$\theta= 15^{\circ},23^{\circ},30^{\circ},45^{\circ},55^{\circ}$ and
energy loss $\nu$. Data are from Refs. \onlinecite{arrd,nicu}.

One notices:

1) For all scattering angles there is good agreement on the elastic side of
the QEP, except for $\theta=23^{\circ}$ where there is a modest
disagreement for the lowest $\nu$. This is similar to the outcome for $^4$He
\cite{viv}, but notably different from all other targets, where  low-$\nu$
predictions fail \cite{rt1,rt11}. The general agreement there may well be due
to the accuracy with which one can calculate $f^{PN,A}$ for the lightest
nuclei \cite{bench} in contrast to targets with $A\ge 12$.

2) The inelastic side of the QEP is usually the one which is best produced for
$A\ge 12$. However, the displayed $D$ predictions reveal discrepancies with
data, in particular for the two lowest angles. Those  get less outspoken for
increasing $\theta$, degenerating in faint wiggles for $\theta=30^{\circ}$. For
$\theta=45^{\circ},55^{\circ}$ the NI part fits very well, but the observed
intensity at, and just beyond  the inelastic side of the QEP,
somewhat exceeds the predictions.

The failure of the underlying picture to describe the above structures is
not inherent to the given description, but is a consequence of the used
parametrization  for $F_k^N=F_k^{N(NI)}$: Those do not account for the
excitation of individual nucleon resonances and instead
averages over those. Their explicit inclusion requires precise information
on transition form factors $G_{M,E}^{N\to \Delta}$, which over the required
$Q^2$-range are not all well enough known. We therefore do not elaborate
on resonance excitations beyond general statements. Yet, qualitatively one
understands from Eq. (\ref{a2}) that $f^{PN,A}$ shifts and broadens resonance
peaks. Only for the $D$ (and then to lowest order), is  the above tantamount
to Fermi broadening. For higher $\theta$ (higher $Q^2$) the QEP and the
resonance peak draw closer, get blurred and are ultimately smoothed out in the
background.

We have still to account for the fact that the QEP and resonance peaks
stand out for the $D$, but not for $A\ge 12$. The reason is the extended
$D$, which causes the normalized $f^{PN}(x,Q^2)$ to attain a much higher
maximum and corresponding narrower width than for an average nucleus. $^4$He
occupies an intermediate position: For $Q^2=3.5\, $GeV$^2$ the  peak values of
$f^{PN,A}(x,Q^2)$ for $D$,$^4$He, $A\ge 12$  are 6.2, 3.1, 1.4-1.6.
The above qualitatively explains the possibility to detect the outstanding
QEP for the $D$ and $^4$He. The data for the latter
for lower $Q^2$ hardly extend beyond the QEP and do barely touch on the
resonance wing.

The fact that the QEP for $D$ and $^4$He are well reproduced by  NE predictions
makes them natural candidates  to study details in the latter and its
potential has been realized in the past for the neutron magnetic form factor
$G_M^n(Q^2)$ \cite{lung}. We have thus finished an analysis of the QEP
parts of the recent $D$ data and of the older NE3 data on $^4$He
\cite{gmn}, constrained by new  information on other static EM form factors
\cite{jones}.

Additional information on charge-current distributions of the neutron,
contained in its Structure Function resides in the inelastic side $x\lesssim 1$
of inclusive cross sections for several, not necessarily light targets.
Somewhere else we shall elaborate on the role
of the $D$ in the extraction of $F_2^n$ \cite{f2nall}.

\section{Conclusion.}

We have computed cross sections for inclusive scattering of 4.045 GeV
electrons on $D$, have discussed general and exceptional features and
have mentioned the potential of the data to obtain information on static
form factors and dynamic structure functions of the neutron. The underlying
theory is precise, but not exact and one should look forward to
improvements, for instance those in the use of a Bethe-Salpeter description
of the $D$ and elastic $p-n$ scattering. Calculation of static form factors
have been completed \cite{ciof2}). An extension to inclusive scattering
would have to go beyond that model, and will somehow have to incorporate
Nucleon SF as for instance in  Eq. (\ref{a2}).

\section{Acknowledgements}

The authors thank John Arrington for having provided tables of the
measured cross sections.

{Figure captions}

Fig. 1 Cross sections for inclusive scattering of 4.045 GeV electrons from $D$
for $\theta=15^{\circ},23^{\circ},30^{\circ},45^{\circ},55^{\circ}$ as function
of the beam energy loss $\nu$. Data are from Ref. \onlinecite{arrd}.


\begin{references}

\bibitem{lung}
A.Lung $et\, al$, Phys. Rev. Lett 70, 718 (1993).

\bibitem{x}
W.P. Schuetz, Phys. Rev. Lett. 38, 259 (1977).

\bibitem{rock}
Rock $et\,al$, Phys. Rev. C26(1982) 1592; I. Sick, D. Day and J.S. McCarthy,
Phys. Rev. Lett. 45, 871 (1980.

\bibitem{day}
D.B. Day $et\, al$, Phys. Rev. C 48 1849 (1993).

\bibitem{arr89}
J. Arrington $et\, al$, Phys. Rev. Lett. 82, 2056 (1999).

\bibitem{arrd}
J. Arrington, private communication.

\bibitem{nicu}
I. Niculescu $et\,al$ Phys. Rev. Lett. 85, 1182 (2000).

\bibitem{gmn}
A.S. Rinat, in preparation.

\bibitem{gr}
S.A. Gurvitz and A.S. Rinat, TR-PR-93-77/ WIS-93/97/Oct-PH; Progress in
Nuclear and Particle Physics, Vol. 34, 245 (1995).

\bibitem{atw}
G.B. West, Ann. of Phys. (NY) 74, 646 (1972); W.B. Atwood and G.B. West,
Phys. Rev. D7, 773 (1973).

\bibitem{pion}
C.H. Llewelyn Smith, Phys. Lett B 128, 107 (1983); M. Ericson and A.W.
Thomas, $ibid$ 112.

\bibitem{rt1}
A.S. Rinat and M.F. Taragin, Phys Rev. C 60, 044601 (1999).

\bibitem{commar}
A.S. Rinat and M.F. Taragin, Phys. Rev. C 62 034602 (2000).

\bibitem{ciof1}
See for instance: C. Ciofi degli Atti, E. Pace and G. Salme, Phys. Rev.
C 43, 1155 (1991).

\bibitem{rt11}
A.S. Rinat and M.F. Taragin, Nucl. Phys. A598, 349  (1996); $ibid$ A620,
A620, 412 (1997); Erratum $ibid$ A623, 773 (1997).

\bibitem{bench}
H. Kamada $et\,al$, Phys. Rev. C 64, 044001 (2001).

\bibitem{gr1}
S.A. Gurvitz and A.S. Rinat, nucl-th/0106032, Phys. Rev. C 65, to be
published.

\bibitem{grs}
H.A. Gersch, L.J. Rodriguez and Phil N. Smith, Phys. Rev.
A5, 1547 (1973).

\bibitem{viv}
M. Viviani, A. Kievsky and A.S. Rinat, nucl-th 0111049, submitted to
Phys. Rev. C.

\bibitem{niko}
A. Bianconi, S. Jeschonnek, N.N. Nikolaev and B.G. Zakharov, Phys. Lett. B
343, 13 (1995).

\bibitem{mach}
R. Machleidt, K. Hohlinde and C. Elster, Phys. Rep. 149, 1 (1987).

\bibitem{rt2}
A.S. Rinat and M.F. Taragin, Nucl. Phys. A 623, 519 (1997).

\bibitem{sag}
S.A. Gurvitz, Phys. Rev. C 42, 2653 (1990).

\bibitem{arn}
P. Amadrauz $et\, al$, Phys. Lett B295, 159 (1992); M. Arneodo $et\, al,\,
ibid$ B364, 107 (1995).

\bibitem{bod}
A. Bodek and J. Ritchie, Phys. Rev. D 23, 1070 (1981).

\bibitem{arn1}
M. Arneodo et al, Phys. Rev. D 50, R1 (1994).

\bibitem{jones}
M. Jones $et\,al$, Phys. Rev. Lett. 84, 1398 (2000); Third Workshop on
'Perspective in Hadronic Physics' Trieste 2001, IT; to be published.

\bibitem{f2nall}
A.S. Rinat and M.F. Taragin, in preparation.

\bibitem{ciof2}
C. Ciofi degli Atti, D. Faralli, A. Yu. Umnikov and L.P. Kaptari,
Phys. Rev. C 60, 034003 (1999).


\end{references}
\end{document}